\documentclass[aip,jcp,10pt,onecolumn,superscriptaddress,balancelastpage,showpacs,preprint]{revtex4-1}
\usepackage[english]{babel}
\usepackage[utf8]{inputenc}
\usepackage[colorinlistoftodos, color=green!40, prependcaption]{todonotes}
\usepackage{centernot}
\usepackage{graphicx}
\usepackage{amsmath}
\usepackage{times}
\usepackage{amssymb}
\usepackage{mathrsfs}
\usepackage{color}
\usepackage{url}
\usepackage{version}
\usepackage[pdftex,colorlinks=true,pdfstartview=FitV,linkcolor= linkcolor,citecolor= linkcolor,urlcolor= linkcolor,hyperindex=true,hyperfigures=false]{hyperref}
\usepackage{balance}
\usepackage{lastpage}
    
\definecolor{linkcolor}{rgb}{0,0,0.6} 

\newcommand{\qqq}{\end{eqnarray}}

\begin{document}

%\title{A statistical-mechanical derivation of the Derjaguin approximation in a molecular solvent}
\title{Derivation of the Derjaguin approximation for the case of inhomogeneous solvents}

\author{H\aa kan Wennerstr\"om}
\email{hakan.wennerstrom@fkem1.lu.se}
\affiliation{Division of Physical Chemistry, Lund University, P.O. Box 124, S-221 00 Lund, Sweden}

\author{Joakim Stenhammar}
\email{joakim.stenhammar@fkem1.lu.se}
\affiliation{Division of Physical Chemistry, Lund University, P.O. Box 124, S-221 00 Lund, Sweden}

\date{\today}

\begin{abstract}
The Derjaguin approximation (DA) relates the force between curved surfaces to the interaction free energy between parallel planes. It is typically derived by considering the direct interaction between the bodies involved, thus treating the effect of an intervening solvent implicitly by a rescaling of the corresponding Hamaker constant. Here, we provide a generalization of DA to the case of a molecular medium between the bodies, as is the case in most applications. The derivation is based on an explicit statistical-mechanical treatment of the contribution to the interaction force from a molecular solvent using a general expression for intermolecular and molecule-surface interactions. Starting from an exact expression for the force, DA is arrived at by a series of well-defined approximations. Our results show that DA remains valid in a molecular solvent as long as (\emph{i}) the surface-molecule interactions are of much shorter range than the radius $R$ of the sphere and (\emph{ii}) the density correlation length in the solvent is smaller than $R$. We then extend our analysis to the case where a phase transition occurs between the surfaces, which cannot easily be covered using a statistical-mechanical formalism due to the discontinuous change in the density of the medium. Instead using a continuum thermodynamic description, we show that this phase transformation induces an attractive force between the bodies, and that the force between curved surfaces can be related to the free energy in the corresponding planar case, in accordance with DA.
\end{abstract}

\maketitle

\section{Introduction} 
\noindent The Derjaguin approximation (DA) relates the force $F(h)$ between two curved surfaces to the interaction free energy per unit area $\mathcal{A}_p(h)$ between two planar surfaces of the same materials,~\cite{Derjaguin:1934,EvansWennerstrom,Israelachvili} and is valid in the limit where the surface-to-surface separation $h$ is much smaller than the radius (or radii) of curvature of the surfaces. Properly applied, it yields significant computational simplifications, and has played a major role in our understanding of the fundamental forces acting in both biological and synthetic colloidal systems, for example via the celebrated DLVO theory of colloidal stability.~\cite{Derjaguin:1941,VerweyOverbeek,Israelachvili,Wennerstrom:2020} Another important practical application of DA is the surface force apparatus technique,~\cite{Israelachvili:2010} which relies on a mapping between the force measured between two crossed cylinders to the interaction free energy between two planar surfaces.

The original statement of the approximation due to Derjaguin~\cite{Derjaguin:1934} is that the interaction force $F(h)$ between two spheres of radii $R_1$ and $R_2$ can be expressed as
\begin{equation}\label{DA_spheres}
F(h) \approx \frac{2\pi R_1 R_2}{R_1 + R_2} \mathcal{A}_p(h),  
\end{equation}
where $h$ is the distance of closest approach between the spheres. The relation for two spheres is easily generalized to other geometries like a sphere and a plane, a plane and a parallel cylinder, or two crossed cylinders: the only resulting modification of Eq.~\eqref{DA_spheres} is the geometrically determined prefactor, while the physical effects are contained in $\mathcal{A}_p$. In Derjaguin's original derivation, as well as in many textbook derivations, $\mathcal{A}_p$ is obtained from the explicit, pairwise summation of a direct interaction potential $u(r)$ acting between the constituent atoms or molecules of the two bodies. Assuming that the entropic contribution to the direct interaction is negligible, this yields an expression for $\mathcal{A}_p(h)$. Equation~\eqref{DA_spheres} then holds under the two conditions that (\emph{i}) $u(r)$ is short-ranged relative to $R_1$ and $R_2$, and (\emph{ii}) $h$ is small relative to $R_1$ and $R_2$. More recently, the validity limits of DA have been experimentally verified,~\cite{Todd:2004,Rentsch:2006,Oversteegen:2003,Oversteegen:2004} and generalizations of DA to more complex geometries, such as anisotropic particles and rough surfaces, have been derived.~\cite{Shen:2012,Torres-Diaz:2017}

For the specific case of van der Waals interactions, $u(r)$ decays as $r^{-6}$ and DA is often derived based on a Hamaker description, where $u(r)$ is explicitly summed to yield $\mathcal{A}_p(h)$ as~\cite{EvansWennerstrom,Israelachvili}
\begin{equation}
\mathcal{A}_p(h) = -\frac{H_{12}}{12 \pi h^2},
\end{equation}
with $H_{12}$ the Hamaker constant describing the interaction between materials 1 and 2. A more sophisticated description of the same physical effect can be obtained using Lifshitz theory,~\cite{Lifshitz:1956,Parsegian} which avoids the assumption of pairwise additivity between interparticle interactions that is particularly poor for the non-dispersion (classical) contributions to the van der Waals interactions. Within the Lifshitz formalism, the two interacting surfaces are treated as continuous materials described by their frequency-dependent dielectric responses. This treatment yields the same distance dependence of the interaction as the Hamaker treatment, although with a modified value of the Hamaker constant $H_{12}$. Furthermore, the effect of a medium between the bodies can be readily included into Lifshitz theory: this medium is also characterized by its bulk dielectric properties, which are assumed independent of the separation between the bodies. In reality, however, the density of the intervening fluid medium will be position-dependent in a way that depends on $h$. For nearly incompressible pure liquids this is a negligible effect, while for a compressible liquid or for media containing more than one component, substantial changes of density or composition can occur as the two bodies approach; an extreme case occurs when a new phase forms between the bodies, as in the case of capillary condensation.~\cite{Petrov:1997} The standard geometrical derivations of DA are not capable of treating such effects of inhomogeneiety, which instead require an explicit statistical-mechanical treatment of the molecular solvent. 

The application of DA to situations with a medium between the surfaces has previously been discussed for a range of specific interactions. Oversteegen and Lekkerkerker~\cite{Oversteegen:2003,Oversteegen:2004} studied DA in the context of the depletion force between hard spherical bodies. Schnitzer and Morozov~\cite{Schnitzer:2015} derived a generalized version of DA valid at any separation for the specific case of electrostatic double layer interactions, while Forsman and Woodward analyzed the validity of DA for two spherical particles in a Lennard-Jones fluid~\cite{Forsman:2010} or in a polymer solution.~\cite{Forsman:2009} In this paper, we provide a complement to these studies, covering the general case of the interaction between curved bodies immersed in a molecular solvent. We present a straightforward, statistical-mechanical derivation of (\emph{i}) an exact expression for the interaction free energy between two infinite planes, and (\emph{ii}) DA for a sphere and a plane, both expressed in terms of the position- and separation-dependent solvent density. Our results show that DA remains valid for the case of a molecular solvent, with two additional constraints compared to the Lifshitz treatment, namely that (\emph{i}) the surface-molecule interactions are of much shorter range than the radius $R$ of the sphere, and (\emph{ii}) the density correlation length in the solvent is smaller than $R$. We then extend our analysis to the case where a phase transition occurs between the surfaces, which cannot be straightforwardly covered using a statistical-mechanical formalism due to the discontinuous change in the density of the medium. Thus, we instead take a continuum thermodynamics approach and show that the phase transformation induces an attractive force between the bodies, whose value between curved surfaces can be related to the corresponding free energy for a planar system in accordance with DA.

\section{The force between a sphere and a plane in a molecular solvent}\label{sec:force_exact}

\begin{figure}[h]
    \center
    \includegraphics[width=12cm]{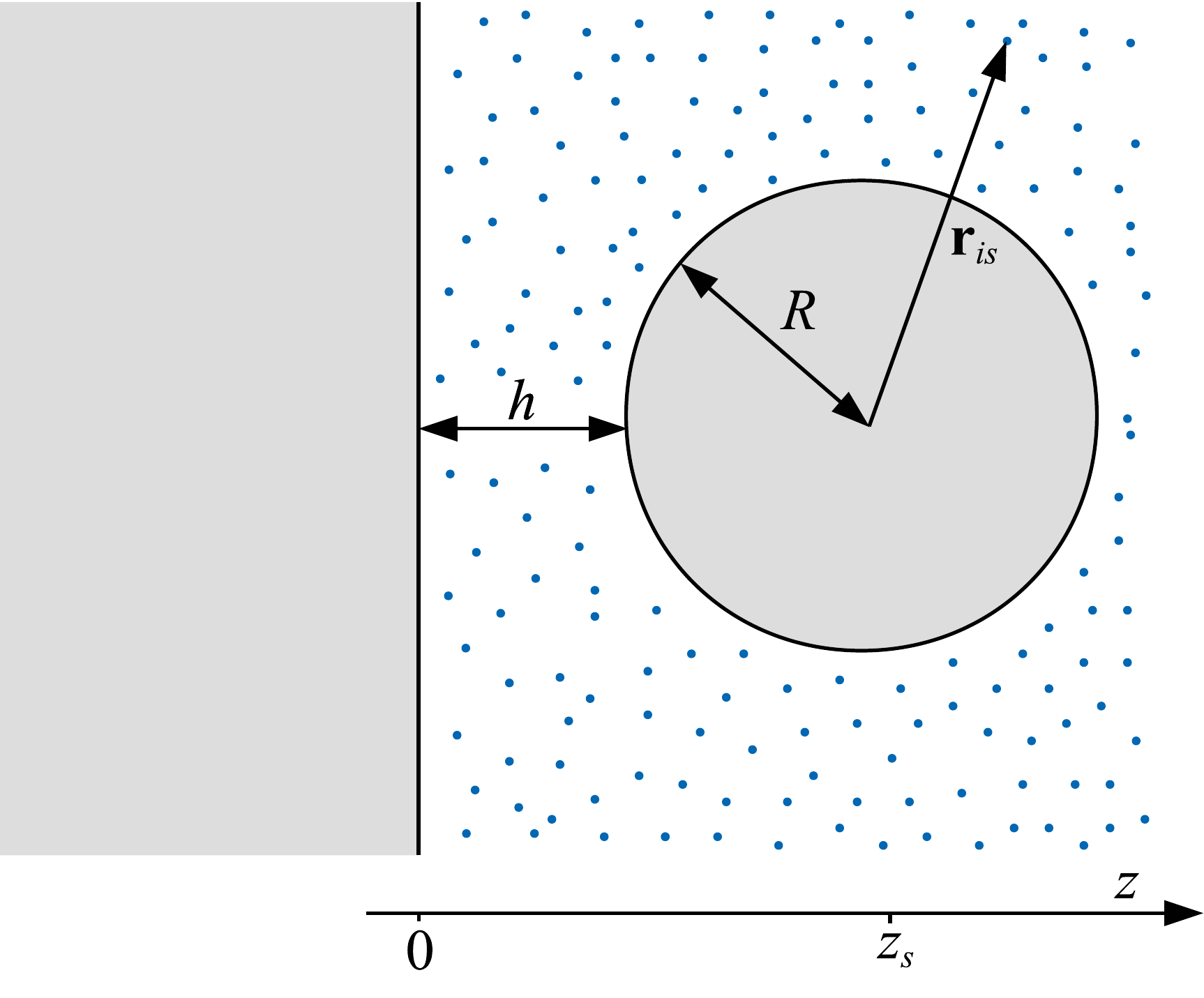}
    \caption{Schematic image of the interaction between an infinite plane and a sphere in a molecular solvent.}\label{fig:1}
\end{figure}

We start by considering the interaction between a homogeneous planar wall placed at $z = 0$ and a sphere of radius $R$ made from the same material as the wall, centered at $\mathbf{r}_s=(0,0,z_s)$, immersed in a molecular solvent containing $N$ molecules in equilibrium with a large reservoir. Molecule $i$ interacts with the wall and the sphere with potentials $u_{iw}(z_i)$ and $u_{is}(r_{is})$, respectively, where $r_{is}$ is the distance between molecule $i$ and the centre of the sphere; see Fig.~\ref{fig:1}. In addition, all molecular pairs $i,j$ interact through a pairwise potential $u_{ij}(r_{ij})$. For simplicity, we consider interactions independent of the molecular orientation; while it is straightforward to generalize the formalism to orientation-dependent interactions, it would lead to a more extensive notation, but adding only marginally to the general understanding. We furthermore ignore the \emph{direct} interaction between the two bodies, which is already covered by the standard derivations~\cite{EvansWennerstrom,Israelachvili} and does not influence the molecular degrees of freedom that are our focus here. The configuration integral $Z_N$ of the solvent particles is formally expressed as
\begin{equation}\label{eq:configuration_int}
Z_N(h) = \int \exp \left[ -\beta U(\{\mathbf{r}_i\}_1^N )\right] \{d\mathbf{r}_i\}_1^N,
\end{equation}
where 
\begin{equation}\label{Utot}
U = \sum_{i=1}^N \left( u_{iw}(z_i) + u_{is}(r_{is}) + \frac{1}{2}\sum_{j \neq i} u_{ij}(r_{ij}) \right)
\end{equation}
is the total energy of the system and $\beta = (k_B T)^{-1}$ the inverse thermal energy. The excess free energy of the system is given by $A(h)=-k_B T \ln Z_N(h)$ and the force between the plane and the sphere is
\begin{equation}\label{F_def}
F(h) = -\frac{dA}{dh} = \frac{k_B T}{Z_N}\frac{dZ_N}{dh}.
\end{equation}
If we vary $h$ by displacing the sphere while keeping the wall fixed, the only term in Eq.~\eqref{Utot} that changes is the one containing $u_{is}$:
\begin{align}\label{Z_derivative} 
& \frac{dZ_N}{dh} = -\frac{1}{k_B T} \int \sum_{i=1}^N  \frac{du_{is}}{dr_{is}} \frac{dr_{is}}{dh} \exp \left[ -\beta U(\{\mathbf{r}_i\}_1^N ) \right]\{d\mathbf{r}_i\}_1^N = \\
& -\frac{N}{k_B T} \int \frac{du_{1s}}{dr_{1s}} \frac{dr_{1s}}{dh} \int \exp \left[ -\beta U(\{\mathbf{r}_i\}_1^N ) \right] \{d\mathbf{r}_i\}_2^N d\mathbf{r}_1,\nonumber 
\end{align}
where the second equality follows from the identity of all solvent molecules. The integral over $\{\mathbf{r}_i\}_2^N$ can be identified as $Z_N\rho(\mathbf{r}_1)$, where $\rho(\mathbf{r}_1)$ is the single-particle density.~\cite{McQuarrie} We furthermore write $r_{1s} = [x_1^2 + y_1^2 + (z_1-R-h)^2]^{1/2}$, so that
\begin{equation}
\frac{dr_{1s}}{dh} = -\frac{1}{r_{1s}}(z_1 - R - h).
\end{equation}
We now change the coordinate system to the center of the sphere, \emph{i.e.}, $z = z_1 - R -h$, which enables us to rewrite Eq.~\eqref{Z_derivative} as
\begin{equation}
\frac{dZ_N}{dh} = \frac{Z_N}{k_B T} \int \frac{du_{s}}{dr} \frac{z}{r} \rho(\mathbf{r}) d\mathbf{r},
\end{equation}
where we have dropped the subscript "1" from the variables. Changing to a spherical coordinate system, performing the trivial integration over $\varphi$ and using Eq.~\eqref{F_def} now yields the following exact expression for the solvent contribution to the force $F(h)$ between a sphere and a plane:
\begin{equation}\label{F_h_exact}
F(h) = 2 \pi \int_0^\pi \int_0^\infty \frac{du_{s}}{dr} \rho(r,\theta) r^2 \sin \theta \cos \theta dr d\theta.
\end{equation}
Using the fact that, as $h \rightarrow \infty$, $\rho$ becomes independent of $\theta$, we can define 
\begin{equation}
\rho_\infty(r) = \lim_{h \rightarrow \infty}\rho(r,\theta).
\end{equation}
Since the force on a free spherical particle in solution vanishes, the integral in Eq.~\eqref{F_h_exact} becomes zero in this limit, we can replace $\rho (r,\theta)$ in Eq.~\eqref{F_h_exact} by $\Delta \rho (r,\theta) \equiv \rho(r,\theta) - \rho_\infty(r)$ to yield
\begin{equation}\label{F_h_deltarho}
F(h) = 2 \pi \int_0^\pi \int_0^\infty \frac{du_{s}}{dr} \Delta \rho(r,\theta) r^2 \sin \theta \cos \theta dr d\theta.
\end{equation}
Note that, while seemingly independent of the wall-molecule interaction $u_w(z)$, the force in Eq.~\eqref{F_h_deltarho} depends implicitly on $u_w$ through its dependence on $\Delta \rho$. A fully analogous derivation of the force per unit area $\mathcal{F}_p(h)$ between two half-planes (corresponding to $R \rightarrow \infty$) furthermore yields
\begin{equation}\label{F_plane_exact}
\mathcal{F}_p(h) = \int_{-\infty}^h \frac{du_w(z)}{dz} \Delta \rho_p(z) dz,
\end{equation}
where $\Delta \rho_p(z)$ is the corresponding density difference between the two planes. Importantly, the integral Eq.~\eqref{F_plane_exact} runs from $z = -\infty$ rather than $z=0$. This is because, even though $\rho_p = 0$ for $z < 0$, $\Delta \rho_p$ is not, since the left wall replaces the solvent as it is brought from infinite to finite separation. This yields a nonzero contribution to the force even for an incompressible solvent between the planes for which $\Delta \rho_p = 0$ in the gap. For the case of van der Waals interactions, this contribution corresponds to the solvent-induced part of the force from the Lifshitz treatment. Equations~\eqref{F_h_deltarho}--\eqref{F_plane_exact} are exact, and will in the following be used to derive DA through a series of approximations. 

\section{Derivation of the Derjaguin approximation}\label{sec:DA}

\begin{figure}[h]
    \center
    \includegraphics[width=15cm]{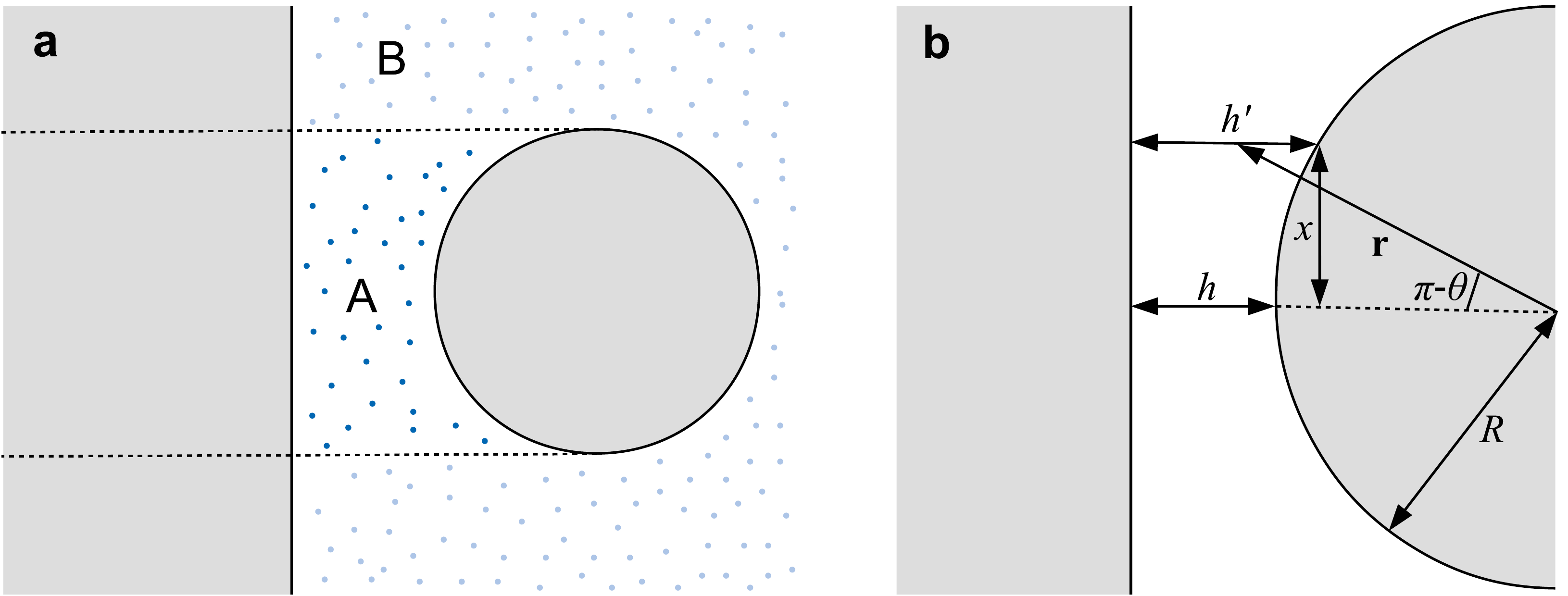}
    \caption{(a) When deriving DA, $\Delta \rho$ is assumed to be zero in region B, so that all contributions to the interaction comes from region A. Note that region A includes the region for $z<0$, for which $\Delta \rho$ is nonzero since the wall replaces the solvent when moved from infinite to finite separations.  (b) Illustration of the coordinate transformation from $\theta$ to $h'$; a direct application of the chord theorem gives $x^2 = r^2 \sin^2 \theta = (h'-h)(2R-h'+h)$.}\label{fig:2}
\end{figure}

We now divide the solvent volume into two parts, as illustrated in Fig.~\ref{fig:2}b: region A directly between the sphere and the wall, and the remainder, region B, outside the projection of the sphere onto the plane. Assuming that the presence of the wall does not affect the density in region B, we can set $\Delta \rho = 0$ here. In region A, we replace $\theta$ by the local perpendicular distance $h'$ between the wall and the sphere surface (see Fig.~\ref{fig:2}b). Using the chord theorem, these variables can be related by
\begin{equation}
\cos^2 \theta = 1 - \frac{(h'-h)(2R-h'+h)}{r^2},
\end{equation}
which, together with Eq.~\eqref{F_h_deltarho}, yields
\begin{equation}\label{F_h_approx1}
F(h) = -2\pi \int_h^{R+h} \int_0^{\infty} \frac{du_{s}}{dr} \Delta \rho(r,h') (R-h'+h) dr dh'.
\end{equation}
Under the additional approximations that (\emph{i}) $h \ll R$, and (\emph{ii}) $\Delta \rho$ is nonzero only in the region where $h' \ll R$; \emph{i.e.}, that the region where $h' \approx R$ does not contribute significantly to the force, we can neglect $h$ and $h'$ in the geometrical factor of Eq.~\eqref{F_h_approx1}, which simplifies to
\begin{equation}\label{F_h_approx2}
F(h) = -2\pi R \int_h^{\infty} \int_0^{\infty} \frac{du_{s}}{dr} \Delta \rho(r,h') dr dh'.
\end{equation}
In Eq.~\eqref{F_h_approx2}, we have extended the upper limit in the outer integral from $R+h$ to $\infty$, again based on the assumption that $\Delta \rho$ is negligible in this region. In order to map the force between a plane and a sphere in Eq.~\eqref{F_h_approx2} to that between two planes, we assume that density correlations in the solvent are of significantly shorter range than $R$, implying that the local density difference $\Delta \rho(r,h')$ between a sphere and a plane is the same as that between two planes at separation $h'$. In other words, we assume that the radius of curvature is large enough that the surface can be regarded as locally flat. Thus, the density variations are unaffected by the local curvature of the surface as long as the density correlation length is smaller than the radius of curvature. Furthermore, we interpret the molecule-sphere force $-\frac{du_{s}}{dr}$ as the limiting value when $R \rightarrow \infty$, so that $\frac{du_{s}}{dr} = \frac{du_{w}}{dz}$ Thus, Eq.~\eqref{F_h_approx2} can be expressed solely in terms of parameters for the planar system:
\begin{equation}
F(h) = -2\pi R \int_h^{\infty} \int_{-\infty}^{h'} \frac{du_{w}}{dz} \Delta \rho_p(z;h') dz dh'.
\end{equation}
From Eq.~\eqref{F_plane_exact}, we can now identify the inner integral as the expression for the force between two planes at separation $h'$, so that 
\begin{equation}\label{Derjaguin1}
F(h) = 2\pi R \int_h^{\infty} \mathcal{F}_p(h') dh'.
\end{equation}
We can now readily identify the integral of $\mathcal{F}_p(h)$ as the free energy $\mathcal{A}_p(h)$, which finally leads us to DA:
\begin{equation}
F(h) \approx 2\pi R \mathcal{A}_p(h).
\end{equation}
In addition to the usual restrictions on $h$ and the range of the direct interaction relative to $R$, we have employed the additional constraints that (\emph{i}) the range of the solvent-sphere interactions is small relative to $R$, and (\emph{ii}) the density correlations in the solvent are short-ranged compared to $R$. As discussed above, this derivation assumes that the interaction between the solvent molecules and the bodies is independent of molecular orientation; including orientational degrees of freedom is straightforward and would lead to position-dependent densities and orientation distributions. This would in turn impose further constraints on the decay of orientational correlations, which need to be short-ranged relative to $R$, leading to complications for a medium close to the isotropic-nematic transition. 

\section{The Derjaguin approximation and capillary-induced phase separation}

Above, we have shown that DA remains valid also for the case when the there are substantial density variations in the medium induced by the surface-surface interactions, under the tacit assumption that the solvent density varies smoothly in space. While all the arguments given above are based on a formally exact expression for the configurational integral, Eq.~\eqref{eq:configuration_int}, complications arise when treating the case of a phase transition taking place between the surfaces, as it is notoriously difficult to treat a first order phase transition using a configurational integral approach. The most important such case is when a capillary induced phase separation occurs as the surfaces approach each other, resulting in an attractive force for both planar and curved surfaces. This practically important case is thus not covered by the formalism of Sections~\ref{sec:force_exact} and~\ref{sec:DA}, and it is not \emph{a priori} obvious whether DA can be used also in this case. One specific complication not present in the case treated above is that a new phase formed between two planar surfaces has an infinite extent in the $xy$ direction, while for curved surfaces, the newly formed phase has a finite volume, which makes a direct application of DA non-trivial. Since we do not have the tools to treat this case using a basic statistical mechanical basis, we instead make use of a continuum thermodynamic approach.

\begin{figure}[h]
    \center
    \includegraphics[width=12cm]{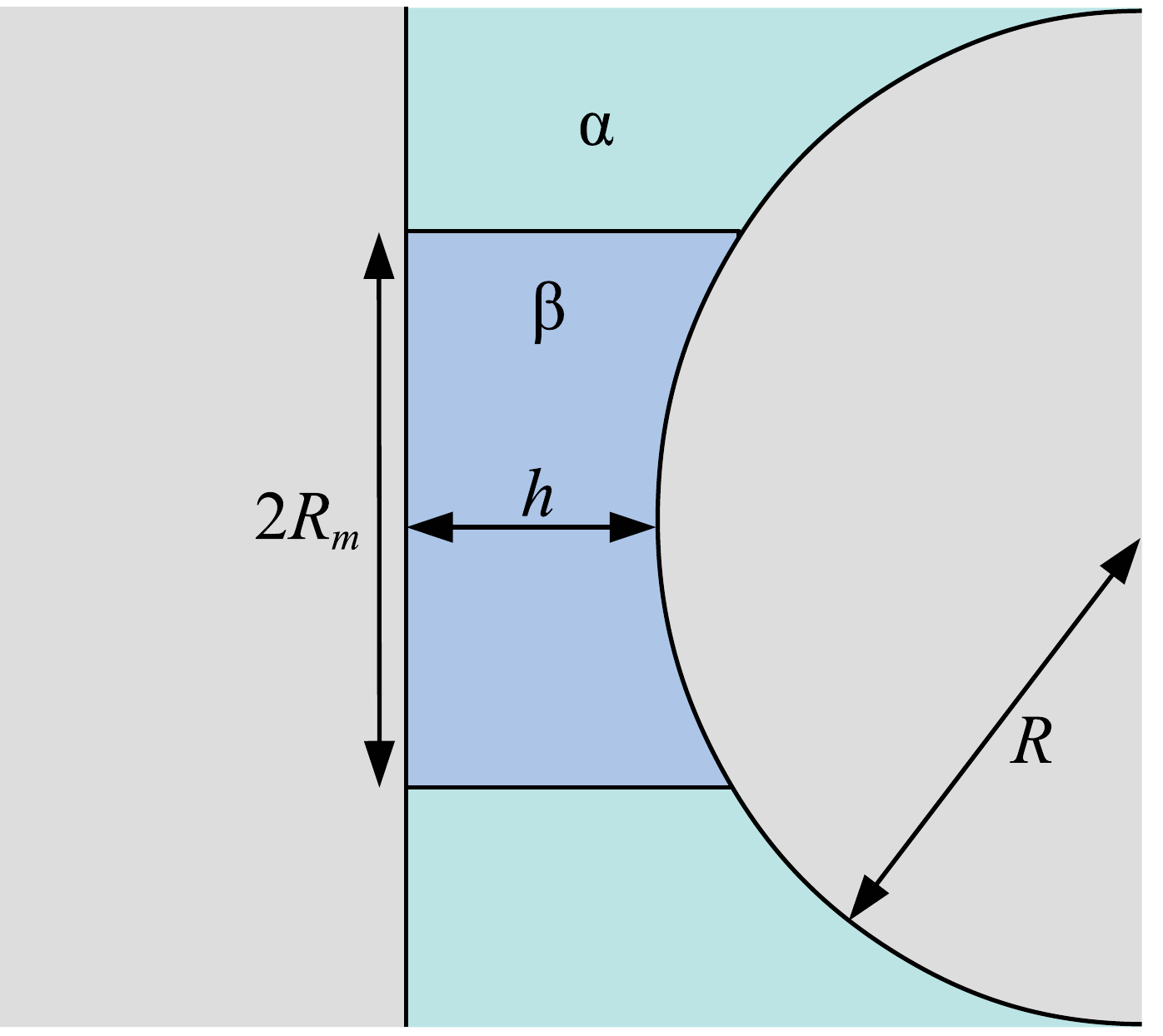}
    \caption{Schematic illustration of a capillary-induced phase transition: A finite lens, approximated as a cylinder, of phase $\beta$ forms in the gap between the two bodies leading to an attractive force.}\label{fig:3}
\end{figure}

Following along the lines of previous derivations,~\cite{EvansWennerstrom,Petrov:1997} we consider an emerging phase $\beta$ that forms a cylindrical lens of radius $R_m$ between a sphere and a plane separated by a minimum distance $h$ immersed in the bulk phase $\alpha$ (see Fig.~\ref{fig:3}). Note that we consider the force between two bodies joined by a medium in equilibrium with an infinite reservoir, so that changes in separation occur at constant chemical potential of the medium, while the case of a condensed phase of constant volume is mathematically much more demanding.~\cite{Lian:2016} The driving force behind the phase transformation is a reduction in surface free energy, quantified by $\Delta \gamma \equiv \gamma_{s\beta}-\gamma_{s\alpha} < 0$, where $\gamma_{s\alpha/\beta}$ is the surface free energy between the solid phase and phase $\alpha/\beta$. The decrease in surface free energy equals $A_s(h) = 2 \pi R_m^2 \Delta \gamma$, while the phase transformation is associated with an increase in bulk free energy, proportional to the volume of the lens and the difference in bulk free energy density $\Delta f = f_\beta - f_\alpha > 0$, and an increase due to the surface energy $\gamma_{\alpha \beta}$ between the two liquid phases. The volume of the lens can be straightforwardly derived by taking into account the volume of the cylinder and that of the spherical cap to second order in $R_m$, yielding $V_{\mathrm{lens}} = \pi R_m^2(h + R_m^2/4R)$. To this order, the total free energy change upon forming the new phase is
\begin{equation}\label{Delta_A_tot}
A(h) = 2\pi R_m^2 \Delta \gamma + \pi R_m^2 \left( h + \frac{R_m^2}{4R} \right) \Delta f + 2 \pi R_m \left(h + \frac{R_m^2}{2R} \right)\gamma_{\alpha \beta}.
\end{equation}
Neglecting, for now, the contribution from the $\alpha \beta$ interface (\emph{i.e.}, putting $\gamma_{\alpha \beta} = 0$), the equilibrium lens radius can be obtained by minimizing Eq.~\eqref{Delta_A_tot} with respecto to $R_m$, yielding
\begin{equation}\label{R_m_eq}
R_m = \sqrt{2R R_K \left( 1-\frac{h}{R_K} \right) },
\end{equation}
where 
\begin{equation}
R_K \equiv -\frac{2\Delta \gamma}{\Delta f}
\end{equation}
is the so-called Kelvin radius. The maximum separation $h_c$ where phase separation occurs can then be calculated by inserting Eq.~\eqref{R_m_eq} into Eq.~\eqref{Delta_A_tot} and setting $A(h) = 0$, yielding 
\begin{equation}
h_c = R_K - \left( \frac{2R_K^2 R_{\alpha \beta}^2}{R} \right)^{1/3},
\end{equation}
where 
\begin{equation}
R_{\alpha \beta} \equiv \frac{2 \gamma_{\alpha \beta}}{\Delta f}.
\end{equation}
This shows that capillary phase separation between a sphere and a plane occurs at a somewhat smaller separation than between two planes, where $h_c = R_K$, due to the surface energy between the two liquid phases. The force $F(h)$ due to the new phase can now be obtained by differentiating Eq.~\eqref{Delta_A_tot}: 
\begin{equation}\label{F_h_capillary}
F(h) = -\frac{dA}{dh} = -2 \pi R R_K \Delta f \left( 1 - \frac{h}{R_K} \right),
\end{equation}
showing that the force is linear and attractive, since $h < R_K$. 

For two parallel planes, the formed $\beta$ phase is infinite, and the free energy per unit area $\mathcal{A}_p(h)$ is readily obtained in an analogous way as Eq.~\eqref{Delta_A_tot}:
\begin{equation}\label{A_plane_capillary}
\mathcal{A}_p(h) = 2 \Delta \gamma + h \Delta f = -R_K \Delta f \left( 1 - \frac{h}{R_K} \right). 
\end{equation}
A direct comparison with Eq.~\eqref{F_h_capillary} shows that 
\begin{equation}
F(h) = 2 \pi R \mathcal{A}_p(h), 
\end{equation}
\emph{i.e.}, that DA holds even in the case of a phase transformation taking place in the solvent, with the additional constraint that $h \ll R_m \ll R$, which was implicitly applied when truncating Eq.~\eqref{Delta_A_tot} at second order in $R_m$. 

\section{Conclusions}

The Derjaguin approximation is a very versatile tool for both the direct interpretation of surface force measurements as well as for the more conceptual understanding of surface forces. In contrast to the standard derivations, based on a direct interaction between two bodies separated by a gap, the derivations presented here consider the validity of DA in the case when the interaction force involves an active participation of a molecular medium surrounding the bodies. Our results show that DA remains generally valid, as is tacitly assumed in many applications, but with two new parameter limitations: First, in analogy with the constraint on the direct interaction between the bodies, the interaction between the solvent molecules and the constituents of the bodies should have a range much shorter than the radius of curvature $R$ of the bodies. The second, somewhat less obvious, constraint is that the density correlation length in the solvent is small relative to $R$, which can be an important limitation in the vicinity of critical points, such as when studying critical Casimir forces.~\cite{Schall:2016} We also showed that DA remains valid even in the case of a new phase forming in the gap, such as for capillary-induced condensation or evaporation, with the additional constraint that the radius $R_m$ of the lens formed by the new phase is smaller than $R$, but larger than $h$. As presented, our derivation covers the case of a single component medium between the bodies. Generalizing our derivation to the case of a two-component system, such as a the case of a molecular solute present in the solvent, where density inhomogeneities are usually more significant, is an interesting route for future work. 

\begin{acknowledgments} 
JS acknowledges funding from the Swedish Research Council (grant IDs 2015-05449 and 2019-03718). Data sharing is not applicable to this article as no new data were created or analyzed in this study. 
\end{acknowledgments}

\newpage
\bibliography{bibliography}
\end{document}